\documentclass[journal]{IEEEtran}
\IEEEoverridecommandlockouts
\usepackage{cite}
\usepackage{amsmath,amssymb,amsfonts}
\usepackage{algorithmic}
\usepackage{subfigure}
\usepackage{graphicx}
\usepackage{textcomp}
\usepackage{color}
\usepackage{multirow}
\ifCLASSINFOpdf
\else
\fi
\begin{document}
\title{Multi-Channel Attentive Feature Fusion for Radio Frequency Fingerprinting}
\author{Yuan Zeng, \emph{Member, IEEE}, Yi Gong, \emph{Senior Member, IEEE}, Jiawei Liu, \emph{Student Member, IEEE}, Shangao Lin, Zidong Han, Ruoxiao Cao, Kaibin Huang, \emph{Fellow, IEEE}, and Khaled B. Letaief, \emph{Fellow, IEEE}
  % <-this % stops a space
  %\thanks{Manuscript received Month Day, Year; revised Month Day, Year; accepted Month Day, Year. Date of publication Month Day, Year; date of current version Month Day, Year.

	\thanks{Y. Zeng is with the Research Institute of Trustworthy Autonomous Systems, Southern University of Science and Technology (SUSTech), Shenzhen, 518055, P. R. China (e-mail: zengy3@sustech.edu.cn).
	
	Y. Gong, S. Lin and Z. Han are with the Department of Electrical and Electronic Engineering, SUSTech, Shenzhen, P. R. China (e-mail: gongy@sustech.edu.cn, linsa2019@mail.sustech.edu.cn, hanzd@sustech.edu.cn).
	
	J. Liu and K. Huang are with the Department of Electrical and Electronic Engineering, The University of Hong Kong, Hong Kong (e-mail: liujw@eee.hku.hk, huangkb@eee.hku.hk).
	
	R. Cao and K. B. Letaief are with the Department of Electronic and Computer Engineering, The Hong Kong University of Science and Technology, Hong Kong (e-mail: rcaoah@connect.ust.hk, eekhaled@ust.hk).}% <-this % stops a space
  % <-this % stops a space
}

%\and

\markboth{}%~Vol.~14, No.~8,
{Shell \MakeLowercase{\textit{et al.}}: Bare Demo of IEEEtran.cls for IEEE Journals}



\maketitle

\begin{abstract}
Radio frequency (RF) fingerprinting is a promising device authentication technique for securing the Internet of Things. It exploits the intrinsic and unique hardware impairments of the transmitters for device identification. Recently, due to the superior performance of deep learning (DL)-based classification models on real-world datasets, DL networks have been explored for RF fingerprinting. Most existing DL-based RF fingerprinting models use a single representation of radio signals as the input, while the multi-channel input model can leverage information from different representations of radio signals and improve the identification accuracy of RF fingerprints. In this work, we propose a multi-channel attentive feature fusion (McAFF) method for RF fingerprinting. It utilizes multi-channel neural features extracted from multiple representations of radio signals, including \textcolor{black}{in-phase and quadrature (IQ) samples}, carrier frequency offsets, fast Fourier transform coefficients and short-time Fourier transform coefficients. The features extracted from different channels are fused adaptively using a shared attention module, where the weights of neural features are learned during the model training. In addition, we design a signal identification module using a convolution-based ResNeXt block to map the fused features to device identities. To evaluate the identification performance of the proposed method, we construct a WiFi dataset, \textcolor{black}{named WiFi Device Identification (WFDI)}, using commercial WiFi end-devices as the transmitters and a Universal Software Radio Peripheral (USRP) platform as the receiver. \textcolor{black}{Experimental results show that the proposed McAFF method significantly outperforms the single-channel-based as well as the existing DL-based RF fingerprinting methods in terms of identification accuracy and robustness.} 
\end{abstract}


% \fi

%\IEEEpeerreviewmaketitle

\section{Introduction}
\label{intro}
With the recent advancement in electronic devices and beyond fifth-generation (B5G) mobile systems, \textcolor{black}{Internet of Things (IoT) technology has been widely developed that plays an increasingly important role in our daily lives, such as smart healthcare, smart cities and intelligent transportation\ \cite{8286847, 8124196, 10101716}. The evolution of IoT tends to enable ubiquitous connectivity, where billions of tiny embedded wireless devices, enabling sensing, computing, and communications, are deployed everywhere. However, wireless security is a critical challenge for IoT applications, since mobile devices in wireless communication networks are vulnerable to malicious attacks when operating in an open environment. Impersonation attack is one of the most common and threatening malicious attacks in wireless communication networks. Device authentication, which validates whether the devices or users are legitimate, can protect the wireless devices from impersonation attacks and improve the security of IoT\ \cite{8377989}.} Traditional authentication methods employed software addresses, such as Internet Protocol (IP) or Media Access Control (MAC), as identity and used classical cryptography-based authentication techniques\ \cite{zou2016survey}. Cryptography algorithms usually rely on complicated mathematical operations or protocols\ \cite{he2014analysis}. However, due to limitations in power consumption and computation resources, such complex cryptography algorithms may be undesirable in many IoT applications, such as smart cities\ \cite{zanella2014internet} and intelligent industries\ \cite{lu2014connected}. An alternative scheme for wireless security is radio frequency (RF) fingerprinting, which uses waveform-level imperfections imposed by the RF circuit to obtain a fingerprint of the wireless device\ \cite{sankhe2019oracle}. The imperfections normally include in-phase and quadrature (IQ) imbalance, phase offset, frequency offset, sampling offset, and phase noise\ \cite{johnson1991physical}, which are usually unique and can hardly be imitated by adversarial devices.

RF fingerprinting is a standard pattern recognition problem and usually consists of two stages: learning and inference. During the learning phase, signals are gathered from end-devices and processed to learn a classification model at a server. After that, the server uses the trained model to authenticate end-devices during the inference phase. Compared with conventional cryptography-based security schemes, RF fingerprinting, by operating at the physical layer, does not impose any additional computational burden and power consumption on end-devices. On the other hand, the receiver/server that is usually equipped with sufficient computational resources is capable of gathering signals from multiple end-devices and performing signal learning and inference. Such arrangement of computational resources and capabilities among the receiver/server and end-devices is particularly desirable for many IoT applications, since most end-devices are low-cost with limited computational and energy resources.

Traditional RF fingerprinting methods are based on carefully hand-crafting specialized feature extractors and classifiers. However, manually extracting handcrafted features requires prior comprehensive knowledge of the communication system, such as channel state and communication protocol. In addition, it is difficult to extract features of hardware imperfection accurately, since the hardware imperfections are interrelated. By performing feature extraction and signal classification at two separate stages, traditional RF fingerprinting methods may work well when the extracted features are distinguishable for classifiers. The identification performance of classifiers is highly dependent on the quality of extracted features, and an efficient feature extraction method for one scenario might easily fail in another.

Recently, deep learning (DL)-based RF fingerprinting methods have been proposed to fingerprint radios through learning of the hardware impairments\ \cite{sankhe2019oracle, riyaz2018deep}. Different neural network architectures, such as convolutional neural networks (CNNs)\ \cite{merchant2018deep}, recurrent neural networks (RNNs)\ \cite{rajendran2018deep} and long short-term memory (LSTM)\ \cite{das2018deep}, have been explored for automatic RF fingerprinting and have shown better RF device identification performance than traditional RF fingerprinting methods. Early DL-based methods mainly used the received IQ samples as the input of the DL network\ \cite{al2020exposing, gong2020unsupervised, pan2019specific}. More recently, a few DL-based RF fingerprinting methods explored the specific characteristics of radio signals, such as bispectrum\ \cite{gong2020unsupervised}, Hilbert spectrum\ \cite{pan2019specific} and differential constellation trace figure\ \cite{peng2019deep}, for RF fingerprinting. \textcolor{black}{The frequency features of radio signals have also been explored for assisting radio signal identification in recent studies\ \cite{zeng2019spectrum, al2020exposing, shen2021radio}, where it is shown that the fast Fourier transform (FFT) and short-time Fourier transform (STFT) coefficients are effective representations of the radio signals, since FFT extracts frequency features of the radio signals and STFT reflects how the instantaneous frequency varies over time.} Utilizing multiple signal representations as the input data may further improve the performance of RF fingerprinting models, since different signal representations can provide different views of the radio signals. In\ \cite{peng2018design}, a hybrid classification framework was proposed to weight multiple features of radio signals with pre-calculated weights for RF fingerprinting. Obtaining an efficient fusion of different representations is a complex task, and pre-calculated weights may produce sub-optimal identification results in practical applications. Therefore, we face the problem of designing an effective learning method to learn multi-channel fusion of different signal representations in an end-to-end manner for better RF device identification. 

In general, training and test data are collected at different time and locations. However, hardware impairments may change across days due to temperature or voltage oscillations, and wireless channels may become unstable due to time-varying fading and noise. Although a lot of progress has been made in RF fingerprinting on radio signals under high signal-to-noise ratio (SNR) conditions, the identification performance still drops sharply when the end-devices are deployed in the presence of electromagnetic interference, environment noises, or other low SNR conditions. A few recent DL-based methods\ \cite{riyaz2018deep, cekic2020robust} proposed model-based augmentation strategies to improve the robustness of the DL-based RF fingerprinting models and validated the proposed strategies using simulated RF devices. These methods provided possible solutions to improving identification accuracy when real RF devices with similar nonlinear characteristics are used. It is still a challenging task to improve the robustness of RF fingerprinting methods in noisy environments.

To tackle the above issues, this paper presents a multi-channel attentive feature fusion (McAFF) model to adaptively fuse multi-channel neural features for RF fingerprinting. \textcolor{black}{We first represent the baseband signal by raw IQ samples, carrier frequency offset (CFO), FFT coefficients, and STFT coefficients. Then, neural features extracted from multi-channel inputs are fused in an end-to-end learning manner. In addition, a shared attention module is designed to learn weights regarding feature importance for RF fingerprinting. The entire model is trained in a supervised manner.}

 \textcolor{black}{Although a few  datasets have been constructed for evaluating DL-based RF fingerprint identification performance in recent studies\ \cite{sankhe2019oracle, al2020exposing, shen2021radio}, they are generally not open-source. In addition, WiFi signals in those datasets were either simulated using software or collected from non-commercial end-devices. To evaluate the performance of the proposed McAFF method for commercial WiFi end-devices, we design a RF signal acquisition system that uses a Universal Software Radio Peripheral (USRP) software defined radio (SDR) platform as the receiver and 72 commercial WiFi end-devices as the target devices.} We collect WiFi signals transmitted using the standard IEEE 802.11n protocol over several experiment days and use the legacy-short training field (L-STF) sequences collected in the uplink frames to construct a dataset for RF fingerprinting, named WiFi Device Identification (WFDI). Extensive experiments are conducted to evaluate the identification performance of the proposed McAFF model. Experimental results demonstrate the effectiveness of the proposed McAFF model on WiFi device identification. The main contributions of our work are summarized as follows.
\begin{itemize}
  \item We propose a novel McAFF model that exploits the complementary information from different representations of radio signals. It optimizes the combination of multi-channel features and utilizes a convolution layer to attend to important features, thus leading to superior RF fingerprinting performance. \textcolor{black}{In contrast to fusing the multiple signal representations using fixed or pre-calculated weights, the proposed McAFF model learns to fuse multiple signal representations adapatively.}
  \item \textcolor{black}{We design a signal identification module that uses a convolution-based ResNeXt block to map the fused neural features to device identities for automatic RF fingerprinting. Moreover, we construct a new dataset with data collected from 72 commercial WiFi end-devices to evaluate the identification performance of the proposed method.}
  \item We provide an extensive ablation study of different signal representations and analyze the rationality of the proposed attention module through comparative experiments. Experimental results demonstrate that the proposed McAFF model outperforms the baseline models in terms of identification accuracy. In addition, we analyze the impact of different data scales and collection dates on the identification performance. Experimental results show that the proposed McAFF model is more robust against noisy radio signals and changes in data scale and collection time.
\end{itemize}

\textcolor{black}{The rest of the paper is organized as follows. Related work is given in Section\ \ref{related}. The problem statement is given in Section\ \ref{pro}. In Section\ \ref{rep}, we briefly introduce the signal pre-processing and representation algorithms. The McAFF model is proposed in Section\ \ref{met}. In Section\ \ref{exp}, experiments are conducted to generate the WFDI dataset and evaluate the WiFi end-device identification performance of the proposed method. Finally, we draw our contribution in Section\ \ref{con}.}

\section{Related Work}
\label{related}
\subsection{RF Fingerprinting}
A variety of approaches have been proposed for RF fingerprinting. Early RF fingerprinting methods usually perform feature extraction and classification separately. Existing hand-tailored feature extraction techniques can be roughly divided into three categories: transient-based (TB) techniques, spectrum-based (SB) techniques, and modulation-based (MB) techniques. TB methods extract the signal variations in the trun-on/off transient, such as the transient signal envelope\ \cite{patel2014improving} and phase offset\ \cite{knox2012practical}. These methods use time domain signal processing algorithms to extract features from received signals.
SB methods extract frequency domain features, such as power spectrum density (PSD)\ \cite{suski2008using, 7122344}, Hilbert spectrum\ \cite{pan2019specific}, and time-frequency statistics\ \cite{bihl2016feature}. MB features including IQ offset\ \cite{brik2008wireless}, clock skew\ \cite{kohno2005remote, jana2009fast}, CFO\ \cite{leonardi2017air, hua2018accurate}, sampling frequency offset\ \cite{vo2016fingerprinting}, etc., can be extracted from the received baseband signal. During the classification process, the authenticator feeds the extracted features to train the classifier and then infers the device identity. Typical classification algorithms include support vector machine (SVM)\ \cite{brik2008wireless} and K-nearest neighbor (KNN)\ \cite{kennedy2008radio}. The identification performance of these classifiers relies heavily on feature extraction, and hand-tailored feature extraction techniques require comprehensive knowledge of communication technologies and protocols, which limits their practical applications.

In the past few years, DL, which utilizes a cascade of multiple layers of nonlinear processing units for feature extraction and transformation, has achieved great success in image classification and speech recognition. Recently, it has also been used for physical-layer communications, including channel estimation\ \cite{lippmann1987introduction, 9018199}, modulation recognition\ \cite{o2016convolutional, zeng2019spectrum, 9336326} and RF fingerprinting\ \cite{soltanieh2020review, shen2021radio, 9913208}. Compared to traditional RF fingerprinting methods, DL-based methods use deep neural networks (DNNs) to automatically extract more distinguishable and high-level features by learning a mapping strategy from training data. A CNN-based RF fingerprinting model, which operates on the error signal obtained after subtracting an estimated ideal signal from frequency-corrected data, was proposed in\ \cite{merchant2018deep}. Riyaz et al.\ \cite{riyaz2018deep} evaluated the effectiveness of an optimized CNN architecture on RF fingerprinting and introduced artificial impairments to improve the identification accuracy. Cekic et al.\ \cite{cekic2020robust} investigated the identification performance of a complex-valued DNN on WiFi and ADS-B signals. Sankhe et al.\ \cite{sankhe2019oracle} used a 16-node USRP SDR testbed and an external database of more than 100 WiFi devices to generate a WiFi dataset and introduced a robust CNN architecture for device identification using raw IQ samples. In\ \cite{sankhe2019no}, an impairment hopping spread spectrum method using a random pseudo-noise binary sequence was proposed. Shen et al.\ \cite{shen2021radio} evaluated the impact of different signal representations, including spectrogram, IQ samples, and FFT coefficients on a CNN-based RF fingerprinting model, and designed a hybrid classifier to improve the RF fingerprint identification performance on LoRa devices. Shawabka et al.\ \cite{al2020exposing} investigated the identification performance of CNN-based RF fingerprinting models on a large-scale dataset with WiFi and ADS-B devices. 

\subsection{Multi-Channel Signal Classification}
Our work is closely related to multi-channel signal classification, which utilizes multiple signal representations to provide classifiers with richer information from different perspectives. Peng et al.\ \cite{peng2018design} proposed combining four modulation features, including differential constellation trace figure, CFO, modulation offset, and IQ offset, with the pre-calculated weights according to the estimated SNR, and they evaluated this feature fusion method on RF fingerprinting using a KNN classifier. In\ \cite{9106397}, a multi-channel learning framework was proposed to exploit the complementary information for automatic modulation recognition. Lin et al.\ \cite{lin2022modulation} introduced a dual-channel spectrum fusion module that includes an original signal channel and a filtered signal channel for modulation recognition. Inspired by existing multi-channel-based classifiers, in this paper, we use four signal representations, namely IQ samples, CFO, FFT coefficients, and STFT coefficients, as the input of our RF fingerprinting model. Moreover, features extracted from these four signal representations are adaptively weighted for RF fingerprinting.

\begin{figure*}[t!]
  \centering
  \includegraphics[scale=0.8]{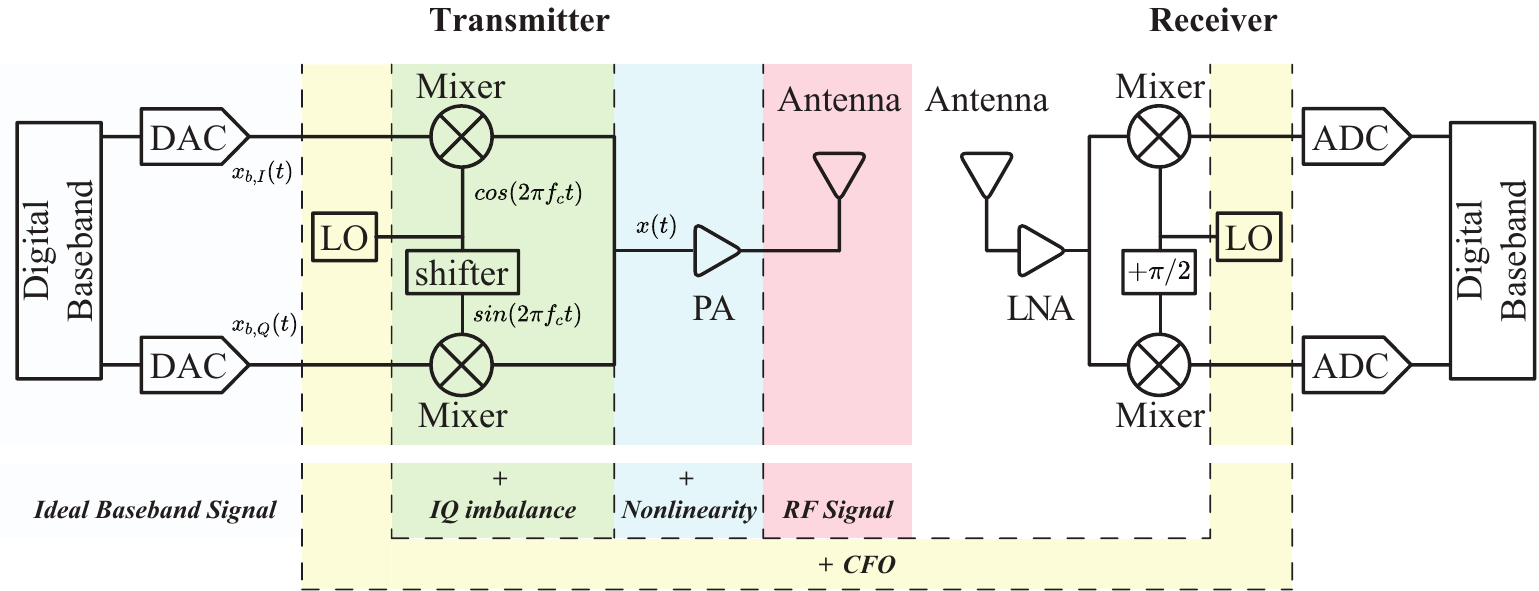}
  \caption{\textcolor{black}{The origins of signal distortions in a communication system.}}
  \label{fig:transmitter_distortion}
\end{figure*}
\subsection{Attention-Based Fusion}
Recently, attention mechanisms\ \cite{vaswani2017attention} that allow the network to automatically find target-related features and enhance them while diminishing other features have achieved great performance improvement on various natural language processing and computer vision tasks\ \cite{moritz2021semi, zhao2020noisy, woo2018cbam}. Fan et al.\ \cite{fan2020spatial} proposed a deep attention-based fusion algorithm to dynamically control the weights of the spatial and spectral features. Xu et al.\ \cite{xu2021attention} presented a co-attention mechanism to guide the fusion of RGB and infrared multi-spectral information for semantic segmentation. In wireless communications, attention mechanisms have been investigated for DL-based modulation classification models to improve classification performance. O'Shea et al.\ \cite{o2016radio} designed an attention mechanism to learn a localization network for blindly synchronizing and normalizing radio signals. Lin et al.\ \cite{lin2022learning} proposed a time-frequency attention mechanism to learn which channel, time, and frequency information is more important for modulation recognition. In this paper, we design a shared attention module to adaptively fuse neural features extracted from multi-channel inputs.

\section{Problem Statement}
\label{pro}
In wireless communication systems, the received signal is typically down-converted to baseband, filtered, and sampled. The complex baseband samples are processed to extract the information bits. The baseband samples are not only degraded by non-ideal channel and noise but also affected by non-idealities of RF and analog parts, including sampling clock offset, power amplifier (PA), phase noise, and CFO. In this section, we introduce several classic RF hardware impairments, including IQ imbalance, CFO, and PA nonlinearity, in the form of mathematical models, as well as the system model of RF fingerprinting.

\subsection{Mathematical Model of RF Impairments}
Let $x_{b}(t)$ denote the transmitted baseband complex signal at time $t$, the in-phase (I) and quadrature (Q) signals of which are denoted by $x_{b,I}(t)$ and $x_{b,Q}(t)$, respectively. $x_{b}(t)$ can be expressed as
\begin{equation}
  x_{b}(t)=x_{b,I}(t)+jx_{b,Q}(t)=r(t)e^{j\phi(t)},
\end{equation}
where $j=\sqrt{-1}$ denotes the imaginary unit. \textcolor{black}{$r(t)$ and $\phi(t)$ denote the amplitude and phase of $x_b(t)$}, respectively. The ideal transmitted RF signal $x(t)$ without hardware impairment is modeled as
\begin{equation}
  x(t)=x_{b,I}(t)\cos(2\pi f_ct)+x_{b,Q}(t)\sin(2\pi f_ct),
\end{equation}
where $f_c$ is the carrier frequency. The RF signal is often distorted in a practical communication system due to the inevitable hardware imperfections, as shown in Fig.\ \ref{fig:transmitter_distortion}, where the IQ imbalance, CFO, and nonlinear distortions of the power amplifier are three typical RF impairments.

\textbf{IQ Imbalance:} In the RF chain of a transmitter, the I and Q baseband signals are up-converted to the cosine and sine carrier waves using two independent quadrature mixers. The cosine and sine carrier waves are generated by the local oscillator (LO) of the transmitter. The ideal sine carrier wave is a copy of the cosine carrier wave delayed by $\pi/2$. However, quadrature mixers are often impaired by gain and phase mismatches between the parallel sections of the RF chain dealing with the I and Q signal paths. The gain mismatch causes amplitude imbalance, and phase deviation from $\pi/2$ results in phase imbalance. \textcolor{black}{Assume that the gains of the I and Q branches are $g_I$ and $g_Q$, respectively, and the phase error is $\varepsilon_p$.} Then the transmitted IQ signals with IQ imbalance can be modeled as
\begin{equation}
\textcolor{black}{ g_I \cos(2\pi f_{c, t}t - \varepsilon_p/2) \ \text{and} \ \ g_Q \sin(2\pi f_{c, t}t + \varepsilon_p/2),}
\end{equation}
where  $f_{c,t}$ is the center frequency that depends on the LO of the transmitter.

\textbf{Non-Linear Distortions of Power Amplifier:} Power amplifiers are used in the RF chain to amplify the pass-band signal for transmission. In practice, the analog amplification process in PA incurs nonlinear distortions in both amplitude and phase. These effects are directly applied to the pass-band signal and can be equivalently modeled on baseband signals. An empirical baseband model of PA was presented in \cite{pedro2005comparative}. \textcolor{black}{Let $g(t)$ and $\epsilon_{\phi}(t)$ denote the amplitude and phase offset of the PA output, respectively}. They can be modeled as
\begin{equation}
  g(t)=\frac{\alpha_a r(t)}{1+\beta_a [r(t)]^2},
\end{equation}
and
\begin{equation}
\textcolor{black}{\epsilon_{\phi}(t) = \frac{\alpha_\phi r(t)^2}{1+\beta_\phi [r(t)]^2},}
\end{equation}
where $\alpha_a, \alpha_\phi, \beta_a $ and $\beta_\phi$ are fitting parameters extracted from measurement results. The equivalent baseband signal with PA nonlinear distortion, denoted by $\tilde{x}_{b}(t)$, is then given by
\begin{equation}
  \tilde{x}_{b}(t)=g(t)e^{j[\phi(t)+\epsilon_{\phi}(t)]}.
\end{equation}

\begin{figure*}[t!]
  \centering
  \includegraphics[scale=1.05]{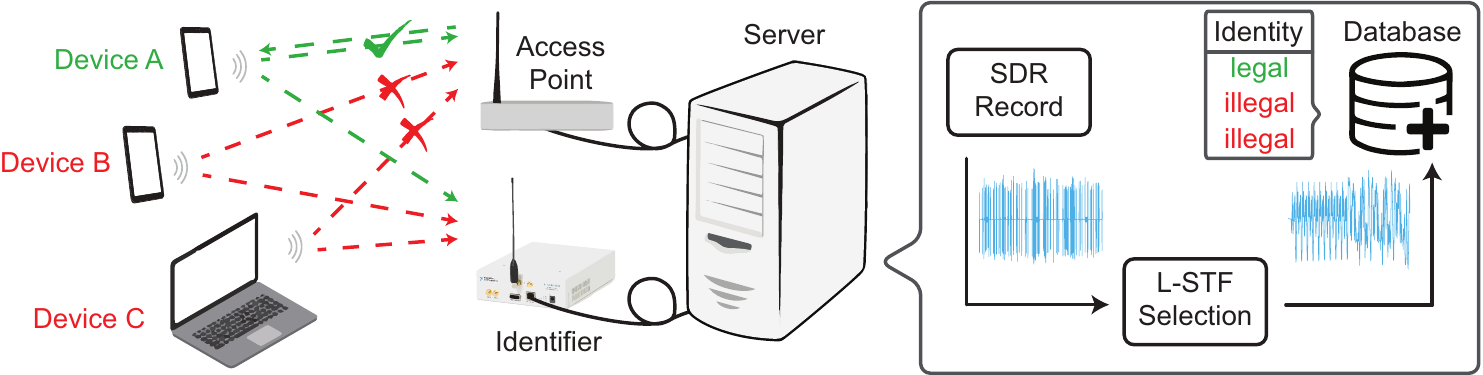}
  \caption{\textcolor{black}{Illustration of a RF fingerprinting system.}}
  \label{fig:system_overview}
\end{figure*}

\textbf{Carrier Frequency Offset:} The carrier frequencies used for up-conversion and down-conversion are generated in the transmitter and receiver using their own LOs, respectively, which can hardly oscillate at an identical frequency. Let $y(t)$ denote the ideal signal received at the receiving antenna. After down-conversion, the received baseband signal is given by
\begin{equation}
  y_{b}(t)=y(t)e^{-j2\pi f_{c,r}t},
	\label{yb}
\end{equation}
where $f_{c,r}$ is the local carrier frequency of the receiver. Let $\epsilon_{f}$ denote the carrier frequency offset between the transmitter and receiver, i.e., $f_{c,r}=f_{c,t}+\epsilon_{f}$. Eq. (\ref{yb}) can be rewritten as
\begin{equation}
  y_{b}(t)=y(t)e^{-j2\pi (f_{c,t}+\epsilon_{f})t},
  \label{baseband_receiver}
\end{equation}
and the carrier waves generated by the receiver's LO can be modeled as
\begin{equation}
  \cos(2\pi (f_{c,t} + \varepsilon_f) t) \ \text{and} \ \  \sin(2\pi (f_{c,t} + \varepsilon_f) t ).
\end{equation}

\subsection{RF Fingerprinting System Model}
\label{RFFing}
RF fingerprinting refers to the task of identifying radio devices by analyzing their transmitted RF signals and extracting RF fingerprints associated with their hardware impairments. \textcolor{black}{An RF fingerprinting system is illustrated in Fig.\ \ref{fig:system_overview}. Specifically, the user device transmits data to the access point (AP) using the standard IEEE 802.11n protocol. The AP is configured to work at the legacy or mixed transmission mode such that the L-STF sequences in the uplink frames are identical. At the same time, the uplink signal is captured by a receiver equipped at the edge server. Throughout this paper, the L-STF sequence is used for signal representation and dataset generation.  
This work aims to identify RF devices using the received baseband signals (corresponding to the L-STF sequences) at the receiver.} Given the received baseband signal $y_{b}(t)$, the server performs a series of signal processing and classification algorithms on $y_{b}(t)$ to identify the RF devices.

\section{Signal Preprocessing and Representation}
\label{rep}

A discrete-time version of the baseband signal $y_b(t)$ in Eq.\ (\ref{baseband_receiver}), denoted by $y(n)$, is obtained by sampling the continuous-time signal with a sampling rate $f_{s}$, associated with a sampling period $T_{s}=1/f_{s}$. That is,
\begin{equation}
  y(n)=y_{b}(t)|_{t=nT_{s}}.
\end{equation}
In preprocessing, we first perform energy normalization on $y(n)$ to remove energy differences between RF devices. Then, we adopt a group of signal representation methods, including CFO estimation, FFT and STFT, to reveal the non-idealities of RF signals.

\subsection{Energy Normalization} The energy of the received signal, despite being used as a feature in conventional classification algorithms, is susceptible to transmission distance and channel conditions. Since RF fingerprinting aims at identifying devices by learning hardware imperfections imposed by the RF circuit, rather than environment differences, we perform energy normalization using the root-mean-square of the amplitude of the signal $y(n)$. The normalized version of $y(n)$ is given by
\begin{equation}
  \tilde{y}(n)=\frac{y(n)}{\sqrt{\frac{1}{N_s}\sum_{n=0}^{N_s-1}y^{2}(n)}},
\end{equation}
where $N_s$ is the number of signal samples. 

\subsection{Signal Representation}

\textcolor{black}{\textbf{IQ Samples:} Let $\tilde{y}_{I}(n)$ and $\tilde{y}_{Q}(n)$ denote the in-phase and quadrature samples of $\tilde{y}(n)$, respectively. Define $\tilde{Y}_I=[\tilde{y}_I(0), \cdots, \tilde{y}_I(N_{s}-1)]^{T}$ and $\tilde{Y}_Q=[\tilde{y}_Q(0), \cdots, \tilde{y}_Q(N_{s}-1)]^{T}$. The IQ samples are concatenated to represent $\tilde{y}(n)$. That is, $\mathbf{R}_{IQ}=[\tilde{Y}_{I}, \tilde{Y}_{Q}]$. The size of $\mathbf{R}_{IQ}$ is $N_{s}\times{2}$.} 

\textbf{CFO Estimation:} CFO is caused by the difference between the LOs of the transmitter and receiver, and results in the phase offset of the received baseband IQ samples, see Eq. (\ref{baseband_receiver}). This work adopts a coarse estimation algorithm introduced in\ \cite{sourour2004frequency} to estimate CFO, that is,
\begin{equation}
  \hat{\epsilon}_{f}=\frac{1}{16} \angle\left(\sum_{n=0}^{N_{s}-17} \tilde{y}^{*}(n) \tilde{y}(n+16)\right),
\end{equation}
where $\angle(\cdot)\in{[-\pi, \pi]}$ denotes the angle of a complex variable, and $(\cdot)^{*}$ denotes the complex conjugation. The phase difference between $\tilde{y}(n)$ and $\tilde{y}(n+16)$ indicates the accumulated CFO over 16 samples. Then, we construct a vector using the phase information of the CFO estimation to represent the signal $\tilde{y}(n)$, that is,
\begin{equation}
  \mathbf{R}_{CFO}=\left[\angle\left(\tilde{y}^{*}(0)\tilde{y}(16)\right), \cdots, \angle\left(\tilde{y}^{*}(N_{s}-17)\tilde{y}(N_{s}-1)\right)\right]^{T}.
\end{equation}
\textcolor{black}{The size of $\mathbf{R}_{CFO}$ is $(N_{s}-16)\times{1}$. After that, we zero-pad $\mathbf{R}_{CFO}$ to expand its size to $N_{s}\times{1}$.}

\textbf{FFT Coefficients:} Given the signal $\tilde{y}(n)$, we use FFT to compute the discrete Fourier transform (DFT) of it, i.e.,
\begin{equation}
  Y_{FFT}(f)=\sum_{n=0}^{N_{s}-1}\tilde{y}(n)e^{-j2\pi fn/N_{s}}, f=0, \cdots, N_{s}-1.
\end{equation}
We use the FFT coefficients, defined by $\mathbf{R}_{FFT}=\left[Y_{FFT}(0), \cdots, Y_{FFT}(N_{s}-1)\right]^{T}$, as a frequency-domain representation of the signal $\tilde{y}(n)$.

\textbf{STFT Coefficients:} STFT is a sequence of DFTs of windowed signals. Unlike FFT where DFT provides the averaged frequency information over the entire signal, STFT computes the time-localized frequency information for situations where frequency components of a signal vary over time\ \cite{KEHTARNAVAZ2008175}. Let $w(\cdot)$ denote a window function of length $J$ and let $K$ be the window shift. The signal $\tilde{y}(n)$ is windowed and transformed into the frequency domain with a DFT, i.e.,
\begin{equation}
  Y_{STFT}(m, f)=\sum_{n=mK}^{mK+J-1}\tilde{y}(n)w(n-mK)e^{-j\omega_{f}(n-mK)},
\end{equation}
where $Y_{STFT}(m, f)$ denotes the STFT coefficients at discrete-time index $m$ and frequency-bin index $f$, and $\omega_{f}=2\pi f/J$ is the discrete frequency variable at frequency-bin index $f$. We use the STFT coefficients $Y_{STFT}(m, f), \forall m, \forall f$, as another representation of the signal $\tilde{y}(n)$, denoted as $\mathbf{R}_{STFT}$.

\section{Feature Fusion and Signal Identification}
\label{met}
\begin{figure}[t!]
  \centering
  \includegraphics[scale=0.21]{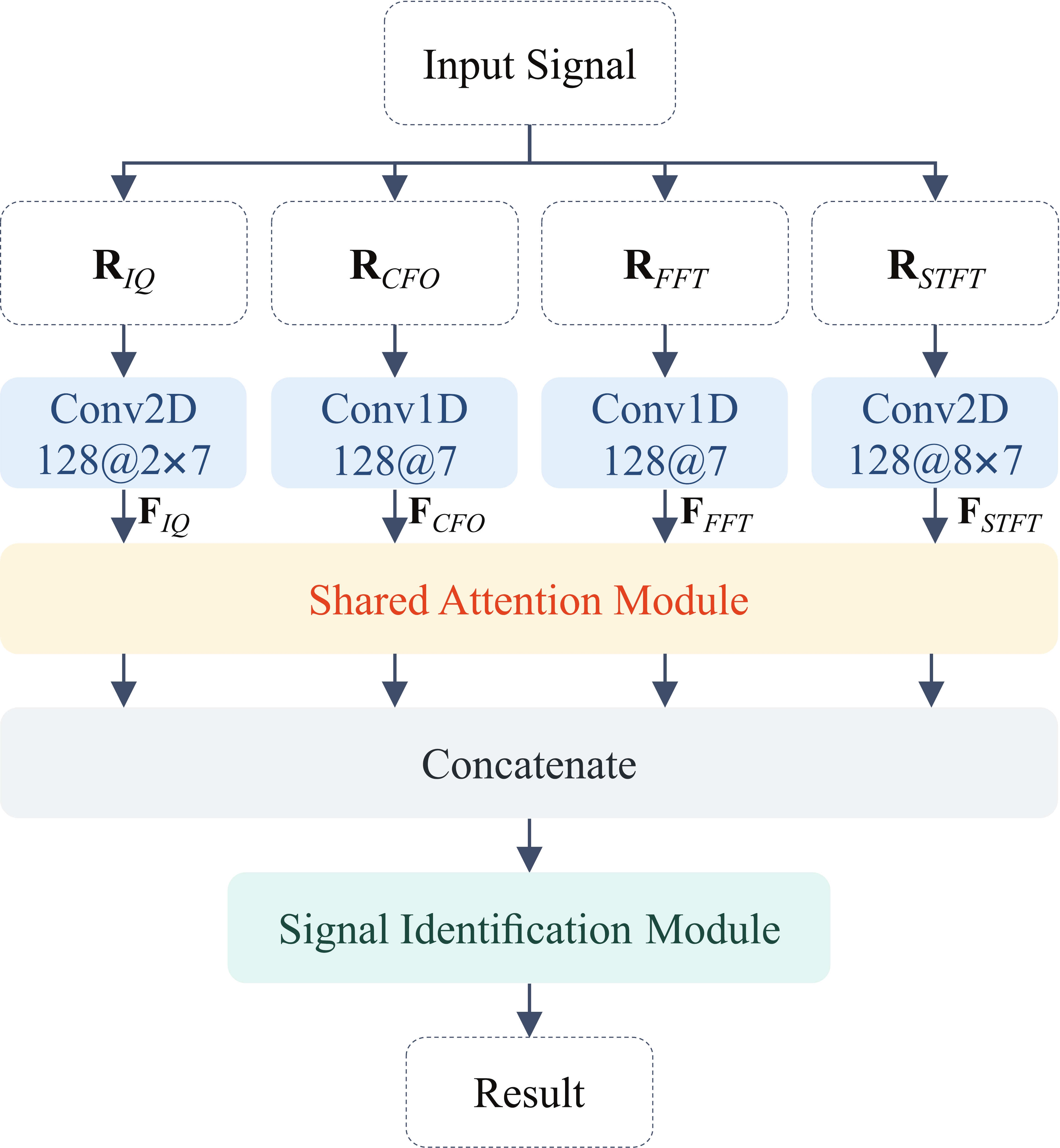}
  \caption{\textcolor{black}{Overview of the proposed McAFF model.}} 
  \label{fig:model}
\end{figure}

This section introduces the proposed McAFF model for RF fingerprinting. The model consists of two functions: multi-channel feature fusion and signal identification. Instead of simply stacking multi-channel inputs including IQ samples, CFO, FFT coefficients and STFT coefficients, a deep attention fusion algorithm is designed to combine them deeply. It uses a shared attention module to adaptively fuse multi-channel neural features, followed by a ResNeXt block\ \cite{xie2017aggregated} based signal identification module. Given the multi-channel inputs, denoted as $\mathbf{R}=\left[\mathbf{R}_{IQ}; \mathbf{R}_{CFO}; \mathbf{R}_{FFT}; \mathbf{R}_{STFT}\right]$, the McAFF model $\mathcal{F}$ is trained to minimize the cross-entropy loss function, that is,
\begin{equation}
 \min \mathbb{E}_{\mathbf{R},\bar{y}}\left(\mathcal{L}(\mathcal{F}(\mathbf{R}), \bar{y})\right),
\end{equation}
where $\mathbb{E}(\cdot)$ denotes the expectation operation and $\mathcal{L}(\mathcal{F}(\mathbf{R}), \bar{y})$ is the cross-entropy loss between the estimation result $\mathcal{F}(\mathbf{R})$ and the true label $\bar{y}$. Fig.\ \ref{fig:model} illustrates an overview of the proposed McAFF model. 

\subsection{Multi-Channel Feature Fusion}
As shown in Fig.\ \ref{fig:model}, the multi-channel inputs  ($\mathbf{R}_{IQ}$, $\mathbf{R}_{CFO}$, $\mathbf{R}_{FFT}$, $\mathbf{R}_{STFT}$) are fed to four separate convolutional layers to acquire four neural feature maps, denoted as $\mathbf{F}_{IQ}$, $\mathbf{F}_{CFO}$, $\mathbf{F}_{FFT}$ and $\mathbf{F}_{STFT}$, respectively. Specifically, the settings of the convolutional layers used to process $\mathbf{R}_{CFO}$ and $\mathbf{R}_{FFT}$ are the same, i.e., one dimensional (1D) convolutional layer with $128$ channels and a filter size of $1\times{7}$. \textcolor{black}{A 2D convolutional layer with $128$ channels and a filter of size $2\times{7}$ is used to process $\mathbf{R}_{IQ}$,} and a 2D convolutional layer with $128$ channels and a filter size  of $8\times{7}$ is used to process $\mathbf{R}_{STFT}$. 

\begin{figure}[t!]
  \centering
  \includegraphics[scale=0.21]{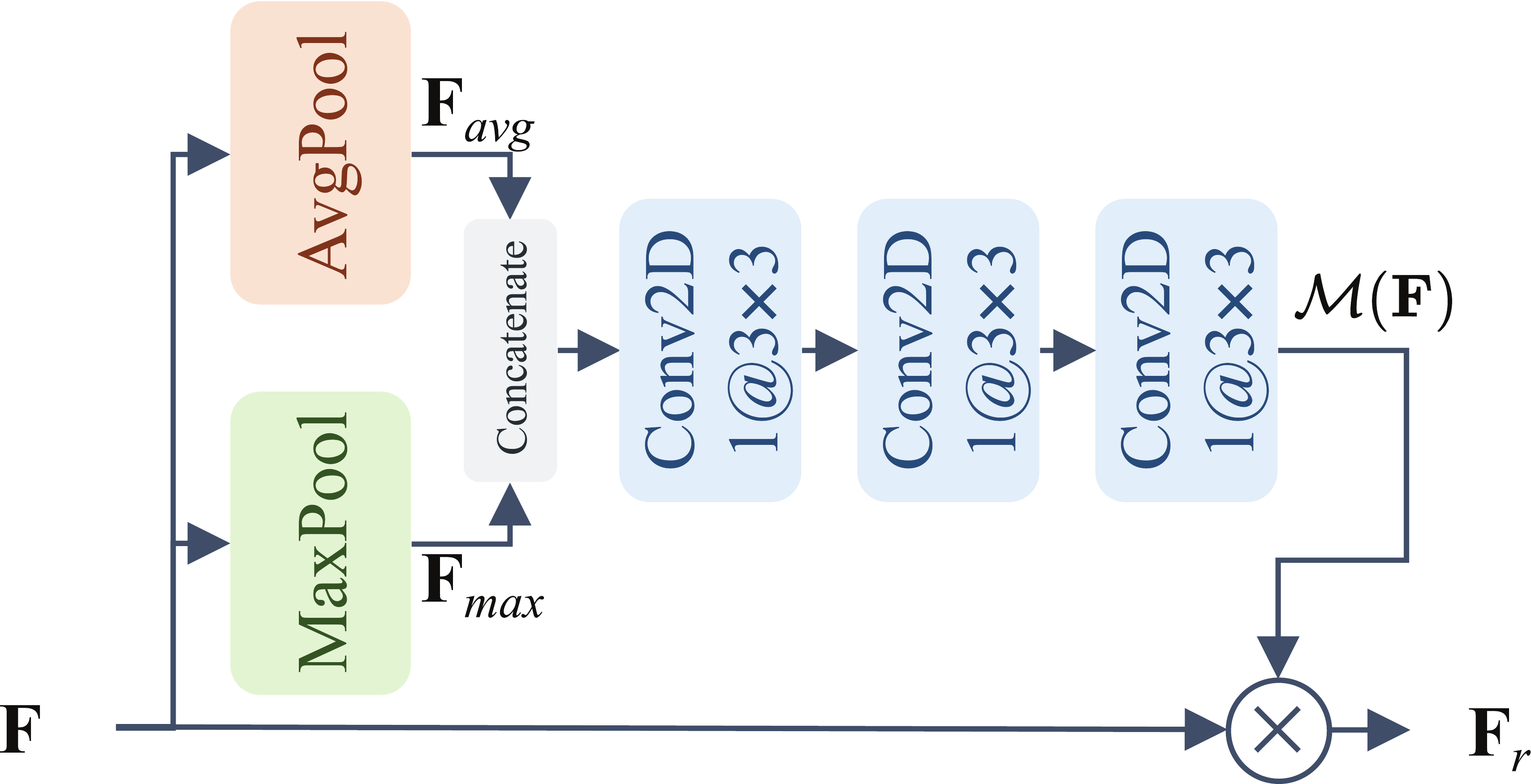}
  \caption{\textcolor{black}{Structure of the proposed shared attention module.}}
  \label{fig:attention}
\end{figure}

The shared attention module is used to refine the intermediate neural feature maps, as illustrated in Fig.\ \ref{fig:attention}. 
Given a feature map $\mathbf{F}\in\mathbb{R}^{C\times{H}\times{W}}$ with $C$ being the number of feature map channels and $H\times{W}$ the size of feature maps, the shared attention module first aggregates spatial information by performing average-pooling and max-pooling in parallel, generating two different spatial context features, denoted by $\mathbf{F}_{avg}\in{\mathbb{R}^{H\times{W}}}$ and $\mathbf{F}_{max}\in{\mathbb{R}^{H\times{W}}}$, respectively. The two context features are then concatenated and convolved by three 2D convolutional layers to infer the attention map \textcolor{black}{$\mathcal{M}(\mathbf{F})\in{\mathbb{R}^{1\times{H}\times{W}}}$}, expressed as
\begin{equation}
  \textcolor{black}{\mathcal{M}(\mathbf{F})=\sigma\left(v_{3}^{3\times{3}}\left(\left[\mathbf{F}_{avg}; \mathbf{F}_{max}\right]\right)\right)},
  \label{Sigmoid}
\end{equation}
where $\sigma$ is the Sigmoid function and $v_{3}^{3\times{3}}$ denotes three cascaded convolutional layers. The attention module $\mathcal{M}$ aims to adaptively control the weight of each point on the input feature map $\mathbf{F}$. Finally, the refined feature map $\mathbf{F}_{r}\in\mathbb{R}^{C\times{H}\times{W}}$ is given by
\begin{equation}
  \textcolor{black}{\mathbf{F}_{r}=\mathcal{M}(\mathbf{F})\otimes\mathbf{F}},
  \label{fine}
\end{equation}
where $\otimes$ is element-wise multiplication. During multiplication, the important parts of the input feature map are enhanced, and the rest ones are faded out according to the learned attention weights $\mathcal{M}(\mathbf{F})$.

\textcolor{black}{With $\mathbf{F}\in\{\mathbf{F}_{IQ}$, $\mathbf{F}_{CFO}$, $\mathbf{F}_{FFT}$,  $\mathbf{F}_{STFT}\}$, the refined versions of the intermediate neural feature maps are denoted as $\mathbf{F}_{r,IQ}$, $\mathbf{F}_{r,CFO}$, $\mathbf{F}_{r,FFT}$, and $\mathbf{F}_{r,STFT}$, respectively. Note that the module $\mathcal{M}$ is shared during training and inference.} After that, the refined neural feature maps are concatenated to generate the fused feature map $\mathbf{F}_{r,c}\in\mathbb{R}^{4C\times{H}\times{W}}$,  given by
\begin{equation}
  \mathbf{F}_{r,c}=\left[\mathbf{F}_{r,IQ}; \mathbf{F}_{r,CFO}; \mathbf{F}_{r,FFT}; \mathbf{F}_{r,STFT}\right].
\end{equation}
The fused feature map $\mathbf{F}_{r,c}$ is then fed to the signal identification module for RF fingerprinting, as shown in Fig.\ \ref{fig:model}.

\subsection{Signal Identification}
\begin{figure}[t!]
  \centering
  \includegraphics[scale=0.20]{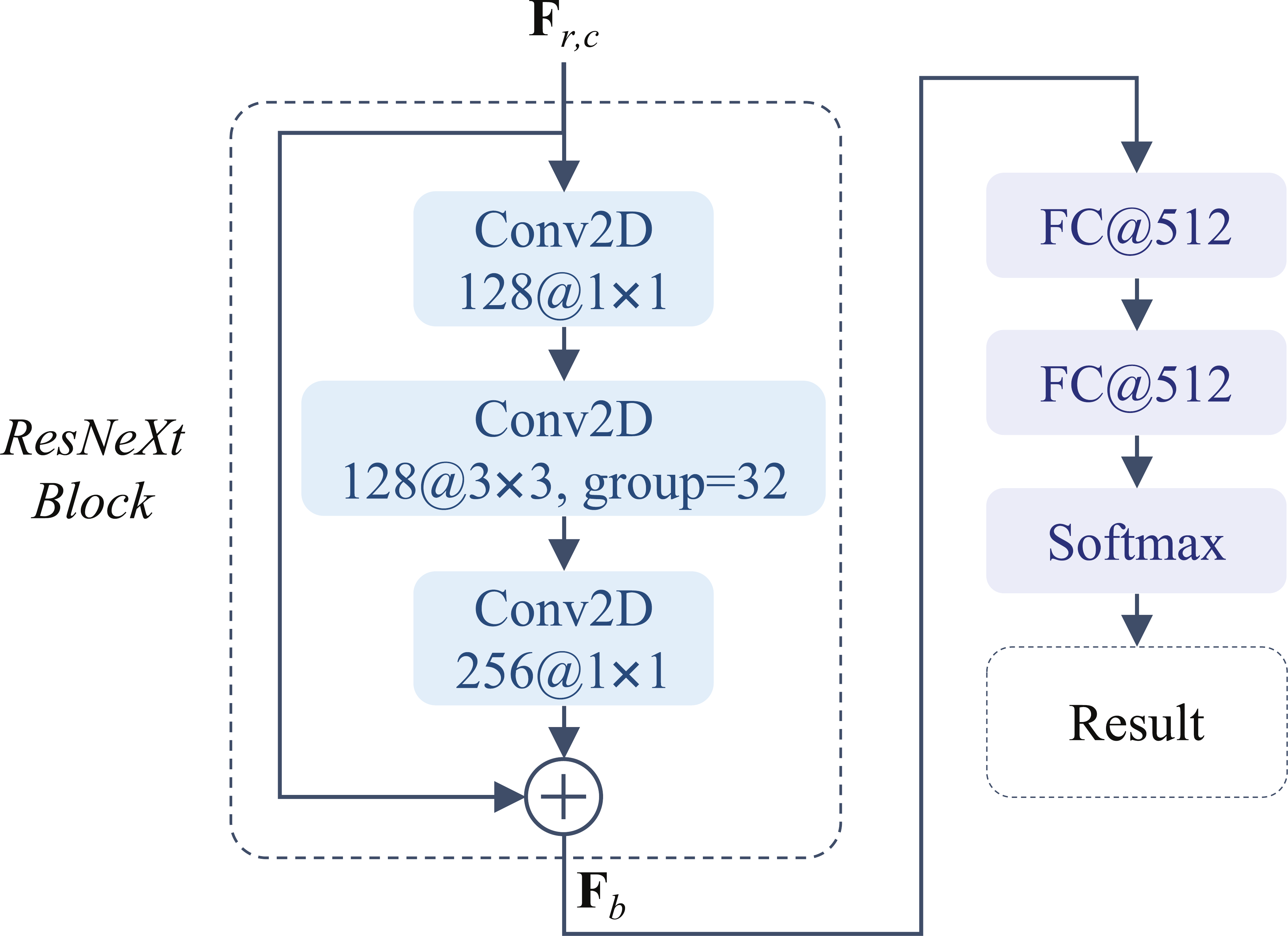}
  \caption{Structure of the proposed signal identification module.}
  \label{fig:sc}
\end{figure}

The signal identification module maps refined features to device identities. As illustrated in Fig.\ \ref{fig:sc}, the signal identification module consists of a convolution-based ResNeXt block\ \cite{xie2017aggregated}, two fully connected (FC) layers and a softmax layer. ResNeXt block is designed to build a simple but highly modularized network architecture for classification. It aggregates a set of transformations with the same topology and exploits the split-transform-merge strategy in an easy and extensible way. Compared with ResNet\ \cite{he2016deep}, ResNeXt block-based neural network is able to improve classification accuracy by increasing the size of the set of transformations. Let $\mathbf{x}=\left[x_{1}, \cdots, x_{D}\right]$ denote a $D$-channel input vector to a neuron and $w_{i}$ denote the weight of a filter for the $i$th channel. A simple neuron in deep neural networks performs the inner product of the input vector and its corresponding weights, which can be thought of as a form of aggregating transformation
\begin{equation}
 \sum_{i=1}^{D}w_{i}x_{i}.
\end{equation}
Such inner product operation is the elementary transformation in fully-connected and convolutional layers. Inspired by the inner production operation, ResNeXt presents a more generic operation by replacing the elementary transformation $w_{i}x_{i}$ with an arbitrary function $\Gamma_{i}(\mathbf{x})$, that is,
\begin{equation}
  \sum_{i=1}^{G}\Gamma_{i}(\mathbf{x}),
  \label{trans}
\end{equation}
where $G$ is the size of the set of transformations to be aggregated. As shown in Fig.\ \ref{fig:sc}, the ResNeXt block processes the input feature map $\mathbf{F}_{r,c}$ and generates the output $\mathbf{F}_{b}$ as 
\begin{equation}
  \mathbf{F}_{b}=\mathbf{F}_{r,c}+\sum_{i=1}^{G}\Gamma_{i}(\mathbf{F}_{r,c}).
\end{equation}
The ResNeXt block consists of three convolution-based operations: a single 2D convolutional layer with 128 channels and filter size of $1\times{1}$, a grouped convolutional layer performing $G=32$ groups of convolutions with 128 channels and filter size of $3\times{3}$, and a single 2D convolutional layer with 256 channels and filter size of $1\times{1}$. The ResNeXt block is followed by two fully-connected layers with 512 neurons each and finally an output layer with the softmax function. Each convolutional and fully-connected layer is followed by batch normalization\ \cite{ioffe2015batch} and Rectified Linear Unit (ReLU)\ \cite{abien2018relu}.

\section{Experiments}
\label{exp}

\subsection{\textcolor{black}{Experimental Setup and Dataset}}
\label{data}
\textcolor{black}{To evaluate the RF fingerprinting performance, we design a signal acquisition system (see Fig. \ref{fig:system_overview}).} The system consists of a wireless AP, WiFi end-devices that transmit WiFi signals to the AP, a National Instrument (NI) USRP-2922 SDR platform that collect WiFi signals from the end-devices, and a server processing the received signals collected at the SDR. The RF signal acquisition system (except the server) is placed in an anechoic chamber, which helps to reduce the impact of the wireless environment. The transmission inside the anechoic chamber is not affected by the external RF activities, and the cones deployed in the anechoic chamber absorb signals generated internally, thus preventing multi-path effects.  %, as shown in Fig.\ \ref{fig:chamber_testbed}
\begin{figure}[t!]
  \centering
  \includegraphics[width=0.9\linewidth]{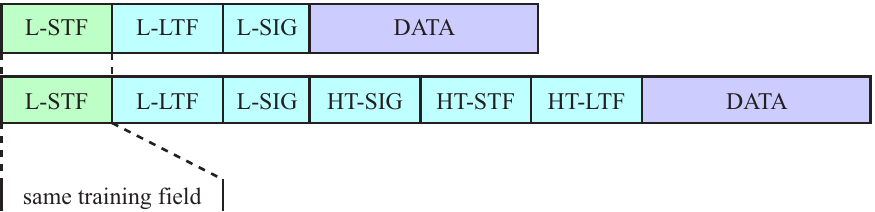}
  \caption{WiFi frame structures at the legacy (upper one) and mixed transmission modes \cite{9363693}.} 
  \label{fig:package_structure}
\end{figure}

\begin{figure}[t!]
  \centering
  \includegraphics[width=0.9\linewidth]{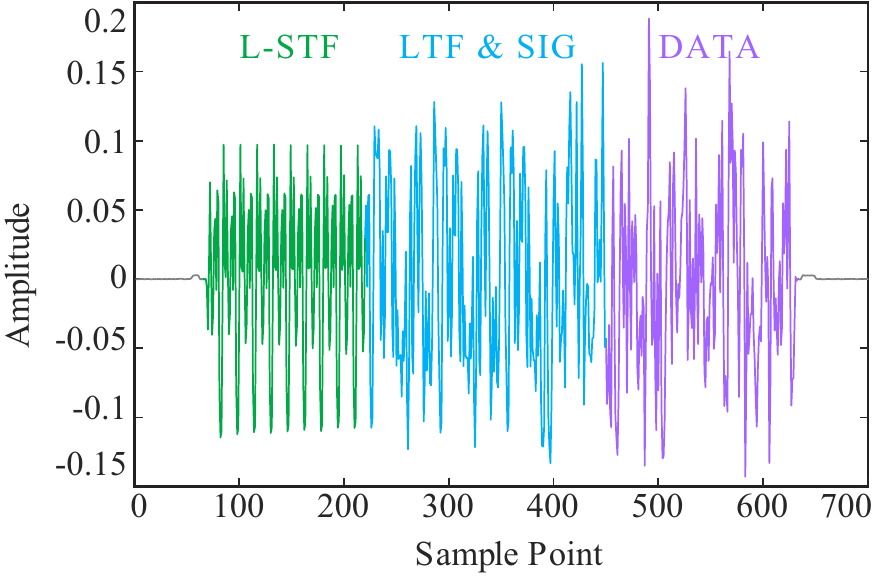}
  \caption{An uplink frame in the legacy transmission mode.} 
  \label{fig:one-second_frame}
\end{figure}

\textcolor{black}{The WFDI dataset is constructed, according to Section IV, by preprocessing the uplink signal of each individual device from the population of 72 commercial WiFi end-devices. 
The communication between the WiFi device and the AP follows the standard IEEE 802.11n  protocol \cite{9363693}. 
The AP is configured to work at the legacy or mixed transmission mode such that the L-STF sequences in the uplink frames sent by different devices are identical. The frame structures at the legacy and mixed transmission modes are illustrated in Fig.~\ref{fig:package_structure}.}

\textcolor{black}{At the SDR receiver, the uplink RF signal sent by the WiFi device is down-converted to baseband and then sampled at a rate of 20 MSa/s for one consecutive second. The one-second signal contains more than 256 uplink frames from the device and approximately the same number of downlink frames from the AP, since the network protocol requires acknowledgement for each frame. In our experiments, the AP is configured to transmit at a lower power than the device, therefore the downlink frames can be easily filtered out by power-based frame detection. An uplink frame collected at the SDR receiver in the legacy transmission mode is illustrated in Fig.~\ref{fig:one-second_frame}. }

\textcolor{black}{In the one-second signal, the standard frame detection algorithm is used to collect $N_f=256$ complete uplink frames. The basic idea is find the peak location of the autocorrelation of L-STF sequences. As prescribed in the IEEE 802.11n protocol, the L-STF sequence is 8 µs long in each frame, which corresponds to 160 samples at the 20 MSa/s sampling rate. In each frame, therefore, the first $N_s=160$ samples are selected to construct the dataset for this device. The rest samples in the frame are discarded since our RF fingerprinting algorithm generates device fingerprint based on the L-STF sequence only. Exploring the utilization of the rest samples in the frame can be considered in our future RF fingerprinting algorithm design.}

\textcolor{black}{The above signal collection experiment is repeated for each device on $N_d=8$ different days at a room temperature of approximately $24^\circ C$.} The WFDI dataset is stored in a matrix form with the dimension of $ N_n \times N_d \times N_f \times N_s \times N_c$, where $N_n=72$ is the number of devices and $N_c=2$ represents the I and Q channels.

\subsection{Implementation Details}

\textcolor{black}{For each device, WDFI consists of $N_{t}=N_d\times{N_{f}}$ frames of data samples and each frame here contains $N_s$ consecutive complex data samples.} To evaluate the identification performance on a training dataset of different sizes, we randomly select a certain percentage of data from the $N_t$ frames. All the experiments are implemented using Keras with Tensorflow backbone and trained on a PC with an NVIDIA TITAN X GPU. The proposed RF fingerprinting model is trained with a batch size of 256 and optimized with the Adam optimizer\ \cite{diederik2015adam}. The learning rate is $0.001$ with a decay factor of $0.1$ and decays when the validation loss does not drop within $10$ epochs. The training is terminated when the validation loss does not drop within 15 epochs. The RF fingerprinting model with the smallest validation loss is saved and used for test. In addition, the window length $J=120$ and window shift $K=5$ are selected for the STFT coefficients calculation. 

\subsection{Experimental Results}

\textcolor{black}{We conduct experiments to evaluate the identification performance of the proposed McAFF model. The other four RF fingerprinting models that use single-channel signal representation for IQ samples, CFO, FFT and STFT coefficients individually are compared with the proposed McAFF model.}

\begin{figure}[t!]
  \centering
	\subfigure[2\%$N_{t}$ training dataset.]{
		\centering
  \includegraphics[width=0.78\linewidth]{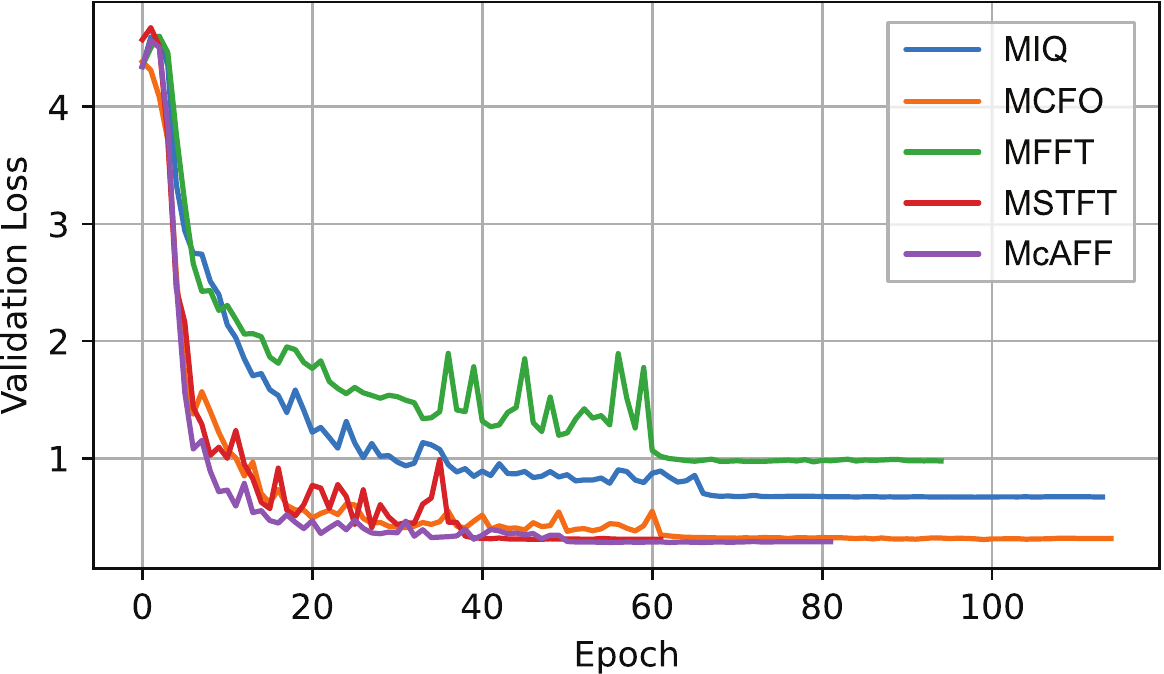}
	\label{2VL}
%	\vspace{1cm}
}
	\subfigure[\textcolor{black}{5\%$N_{t}$ training dataset.}]{
		\centering
  \includegraphics[width=0.80\linewidth]{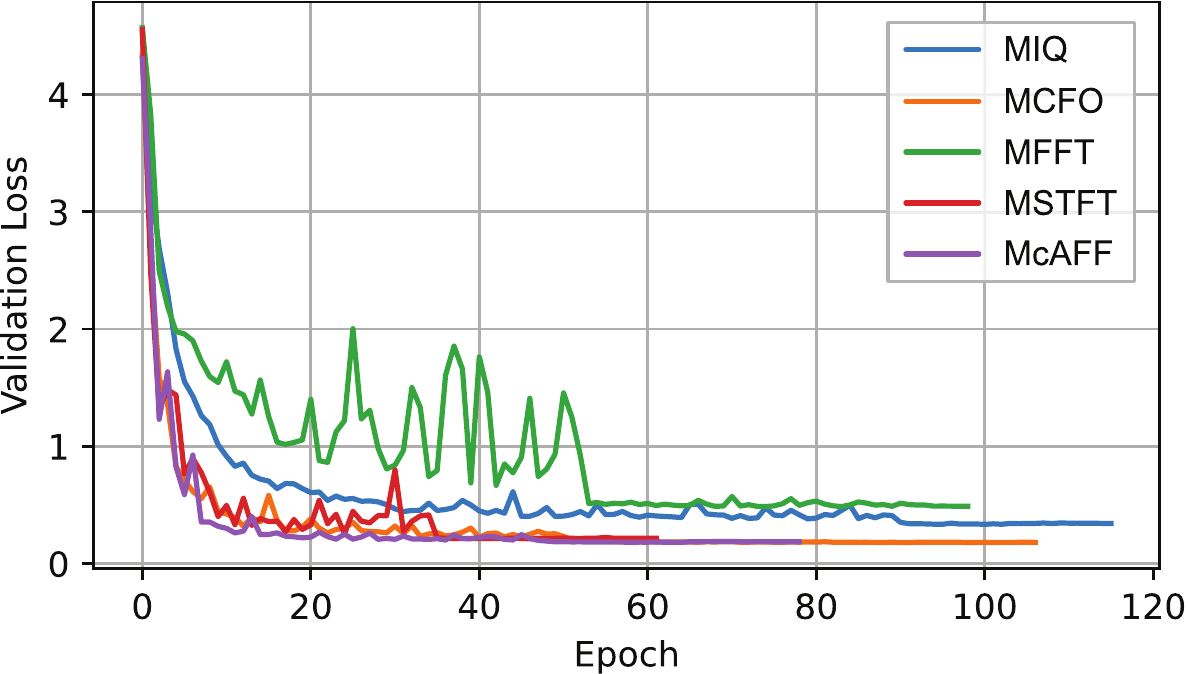}
	\label{5VL}
	}
  \caption{\textcolor{black}{Validation loss versus training epochs.}}
  \label{fig:training_curve}
\end{figure}

\textbf{Ablation study of signal representations:} We do ablation studies to illustrate the impact of different signal representation on RF device identification accuracy. 
\textcolor{black}{The validation losses of different RF fingerprinting models are shown in Fig.~\ref{fig:training_curve}, where the RF fingerprinting models using single-channel signal representation for IQ samples, CFO, FFT, and STFT coefficients are referred to as MIQ, MCFO, MFFT, and MSTFT, respectively. It shows how the validation cross-entropy loss varies when increasing the number of training epochs. Compared with the single-channel RF fingerprinting models, i.e., MIQ, MCFO, MFFT and MSTFT, the proposed McAFF model not only converges faster, but also converges  to a lower validation loss. We also observe that all the models converge to a lower validation loss when the training dataset size increases from 2\%$N_{t}$ to 5\%$N_{t}$. Table\ \ref{table:data_size_vs_acc} further shows the test performance of RF fingerprinting for different training dataset sizes.} When the training dataset size is 50\%$N_{t}$, all the models can provide excellent accuracies (over 96\%) for device identification. The proposed McAFF model outperforms all the other models under the same training dataset sizes. It is also more robust to the change of training dataset sizes. 

\begin{table}[t!]
  \renewcommand{\arraystretch}{1.3}
  \caption{\textcolor{black}{RF fingerprint identification performance comparison.}} %
  \begin{center}
    \begin{tabular}{c|cccccccc}
      \hline \hline
               & 1\%$N_{t}$                           & 5\%$N_{t}$              & 15\%$N_{t}$              & 25\%$N_{t}$              & 50\%$N_{t}$              \\
      \hline
      MIQ      & 64.52\%                    & 91.17\%          & 95.95\%          & 97.21\%          & 98.29\%          \\
      MCFO     & 86.69\%                   & 96.06\%          & 97.92\%          & 98.38\%          & 98.94\%          \\
      MFFT     & 47.37\%                   & 86.43\%          & 93.27\%          & 94.3\%           & 96.17\%          \\
      MSTFT    & 89.94\%                  & 95.18\%          & 96.81\%          & 97.41\%          & 98.01\%          \\
      McAFF & \textbf{90.00\%}  & \textbf{96.11\%} & \textbf{97.94\%} & \textbf{98.41\%} & \textbf{99.01\%} \\
      \hline \hline
    \end{tabular}
    \label{table:data_size_vs_acc}
  \end{center}
\end{table}

\textbf{``Train and Test Same Days" scenario:} We split our WFDI dataset into 8 sub-datasets according to the collection date and name the sub-dataset collected on day $i$ as $D_{i}, i=1,2,\cdots,8$. 
We train the McAFF model on each sub-dataset and get a model for each day. Table\ \ref{table:same_day_acc} shows the identification results of the proposed method on different sub-datasets. It is observed that the proposed McAFF model works pretty well when training data and test data are collected on the same day. Specifically, the proposed McAFF model achieves the highest identification accuracy of 99.85\% on $D_{8}$ and the lowest one of 98.21\% on $D_{7}$. The average identification accuracy of the proposed McAFF model over eight days is 99.12\%. 

\begin{table}[t!]
  \setlength{\tabcolsep}{1.25mm}%\renewcommand{\arraystretch}{1.3}
  \caption{RF fingerprint identification performance of the proposed McAFF model in the ``Train and Test Same Days" scenario.}
  \begin{center}
    \begin{tabular}{cccccccc}%c|
      \hline \hline
                $D_{1}$ & $D_{2}$ & $D_{3}$ & $D_{4}$ & $D_{5}$ & $D_{6}$ & $D_{7}$ & $D_{8}$ \\
      \hline
       99.17\%   & 98.25\%   & 99.26\%   & 99.62\%   & 99.01\%   & 99.6\%    & 98.21\%   & 99.87\%   \\%McAFF &
      \hline \hline
    \end{tabular}
    \label{table:same_day_acc}
  \end{center}
\end{table}

\begin{table}[t!]
  \setlength{\tabcolsep}{1.25mm}%\renewcommand{\arraystretch}{1.3}
  \caption{RF fingerprint identification performance comparison in the ``Train and Test Different days" scenario.}
  \begin{center}
    \begin{tabular}{c|cccc}
      \hline \hline
               & 1\_train/7\_test & 2\_train/6\_test         & 3\_train/5\_test         & 4\_train/4\_test         \\
      \hline
      MIQ      & 58.78\%  & 75.81\%          & 79.76\%          & 82.81\%          \\
      MCFO     & 62.44\%  & 75.26\%          & 78.9\%           & 81.18\%          \\
      MFFT     & 70.22\%  & 81.39\%          & 84.58\%          & 86.45\%          \\
      MSTFT    & \textbf{76.10\%}  & 85.43\%          & 86.52\%          & 88.06\%          \\

      McAFF & 75.35\%  & \textbf{86.04\%} & \textbf{87.72\%} & \textbf{88.67\%} \\
      \hline \hline
    \end{tabular}
    \label{table:diff_multi_day_acc}
  \end{center}
\end{table} 

\textbf{``Train and Test Different Days" scenario:} In this experiment, the training data and test data are collected on different days. Table\ \ref{table:diff_multi_day_acc} shows the experimental results. When data collected from 1 day is used for training and data collected from the remaining 7 days is used for test (referred to as 1\_train/7\_test in Table\ \ref{table:diff_multi_day_acc}), the proposed method has slightly worse identification accuracy than MSTFT. It is observed that by using more training data collected from more days (the other data from the remaining days is used for test), all the models achieve better performance and the identification accuracies are improved significantly. In general, the proposed McAFF method still achieves the best identification performance. This is not unexpected since the proposed method uses multi-channel inputs and adaptively fuses the multi-channel neural features. Therefore, it is more robust to different data collection dates.

\begin{figure}[t!]
  \centering
  \includegraphics[width=\linewidth]{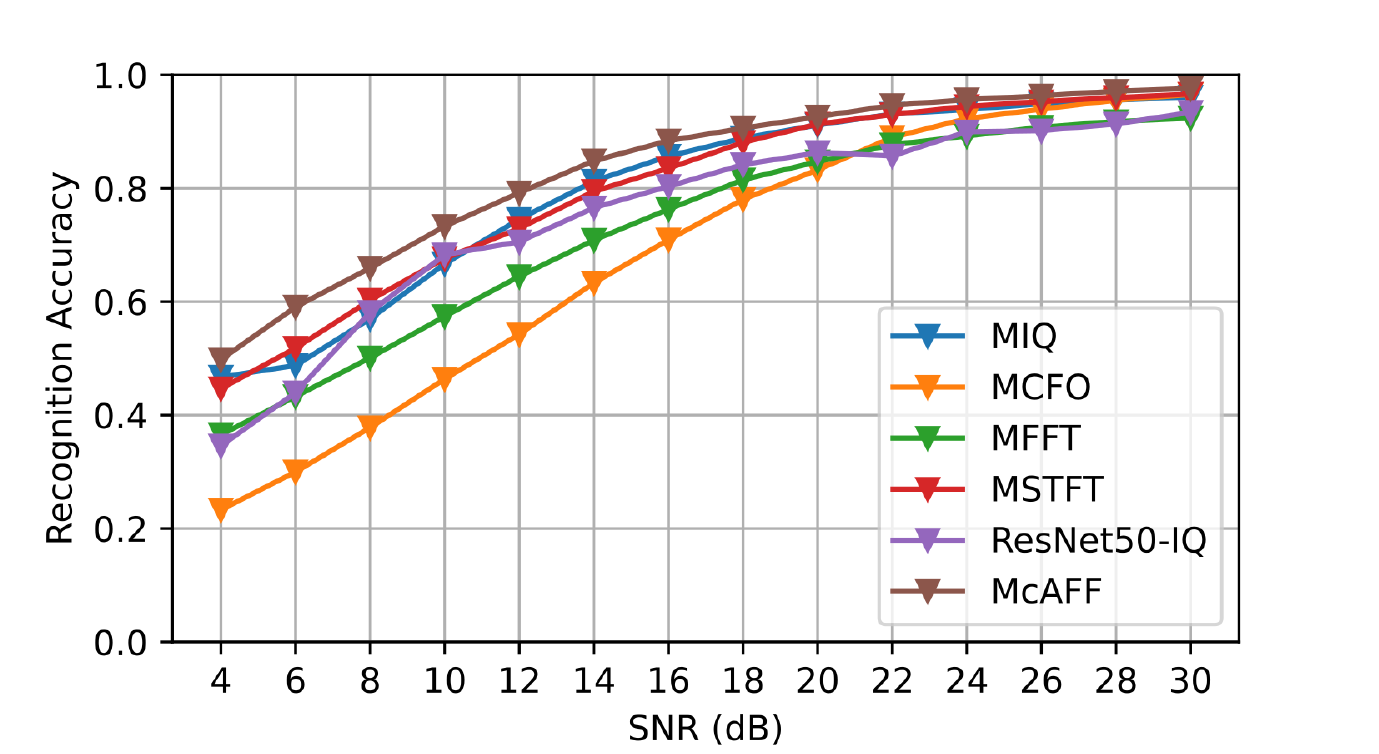}
  \caption{\textcolor{black}{RF fingerprint identification performance versus SNR. The training dataset size is $50\%N_{t}$.}}
  \label{fig:noise}
\end{figure}

\begin{figure*}[thb]
\centering
\subfigure[1\_train/7\_test.]{
\begin{minipage}[t]{0.5\linewidth}
\centering
\includegraphics[width=3.5in]{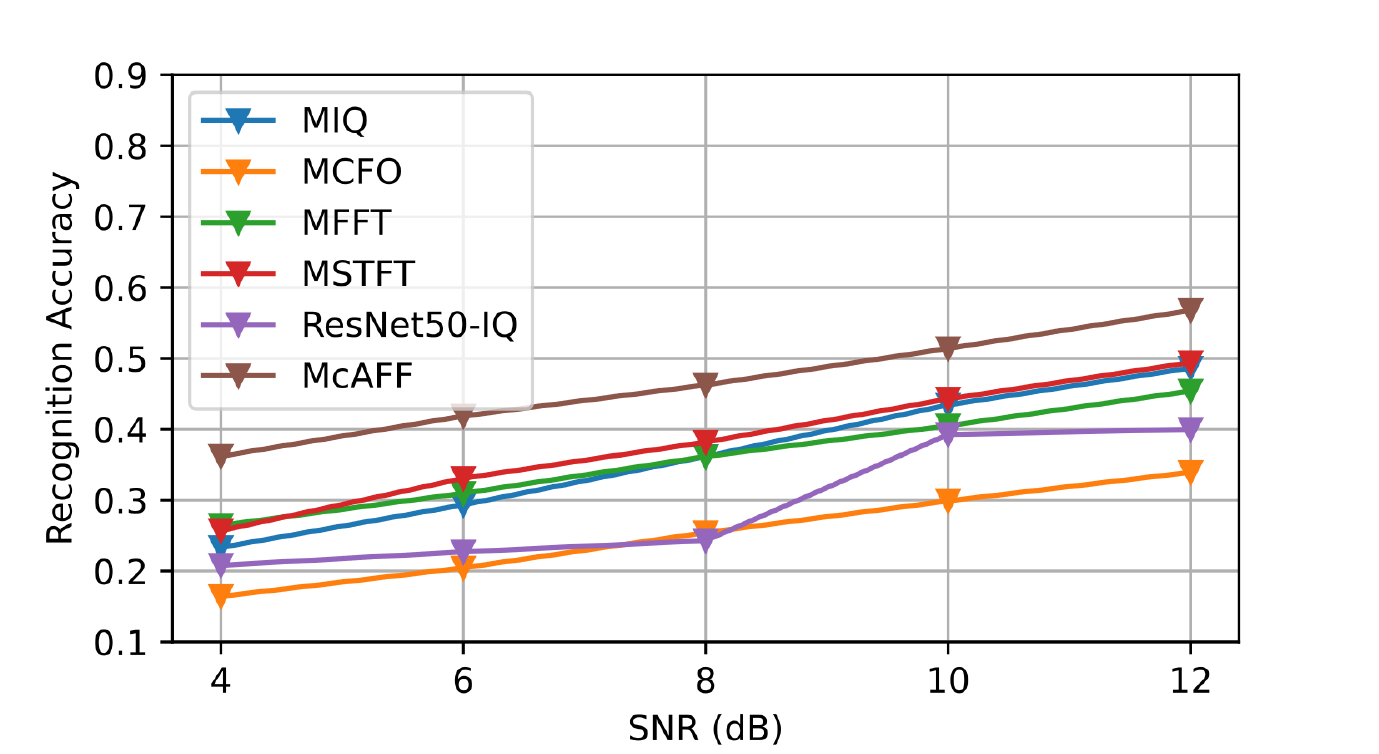}
%\caption{fig2}
\end{minipage}
}%
\subfigure[2\_train/6\_test.]{
\begin{minipage}[t]{0.5 \linewidth}
\centering
\includegraphics[width=3.5in]{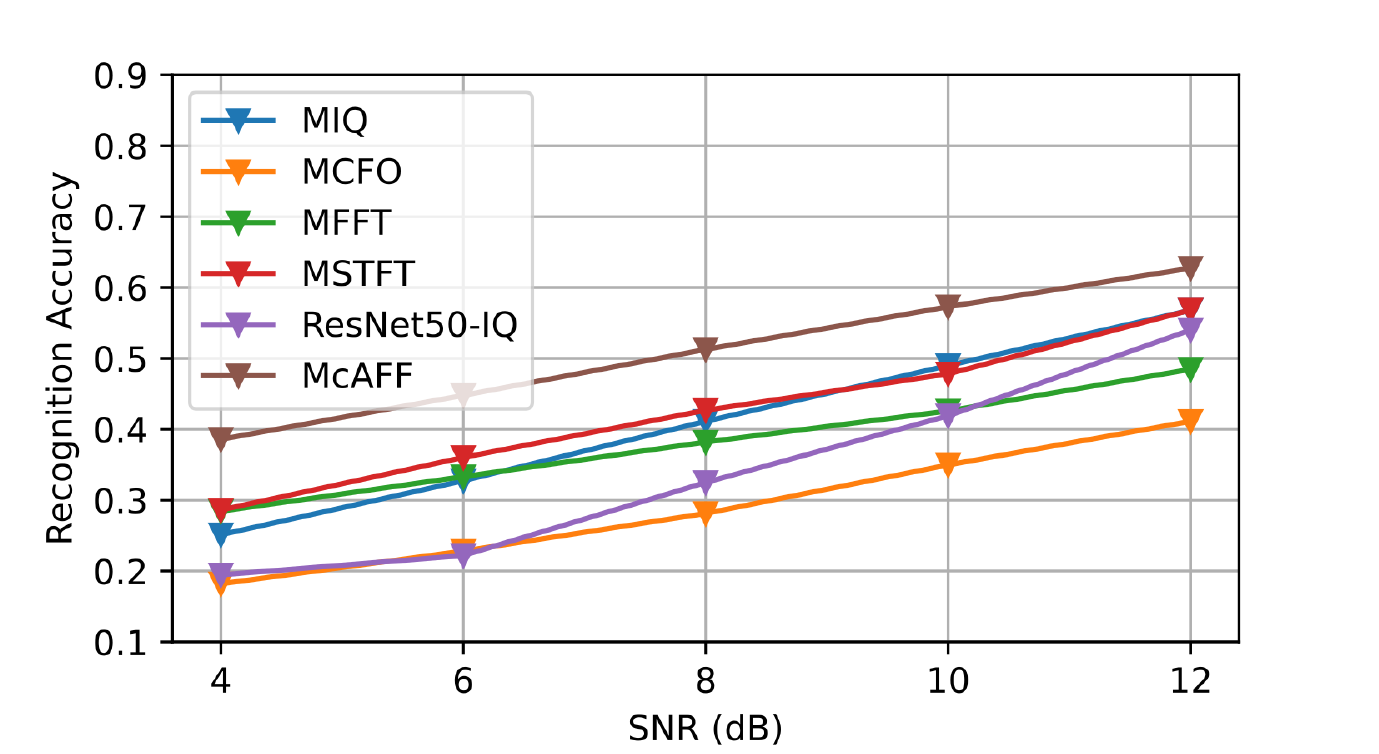}
%\caption{fig2}
\end{minipage}
}%
\centering

\subfigure[3\_train/5\_test.]{
\begin{minipage}[t]{0.5\linewidth}
\centering
\includegraphics[width=3.5in]{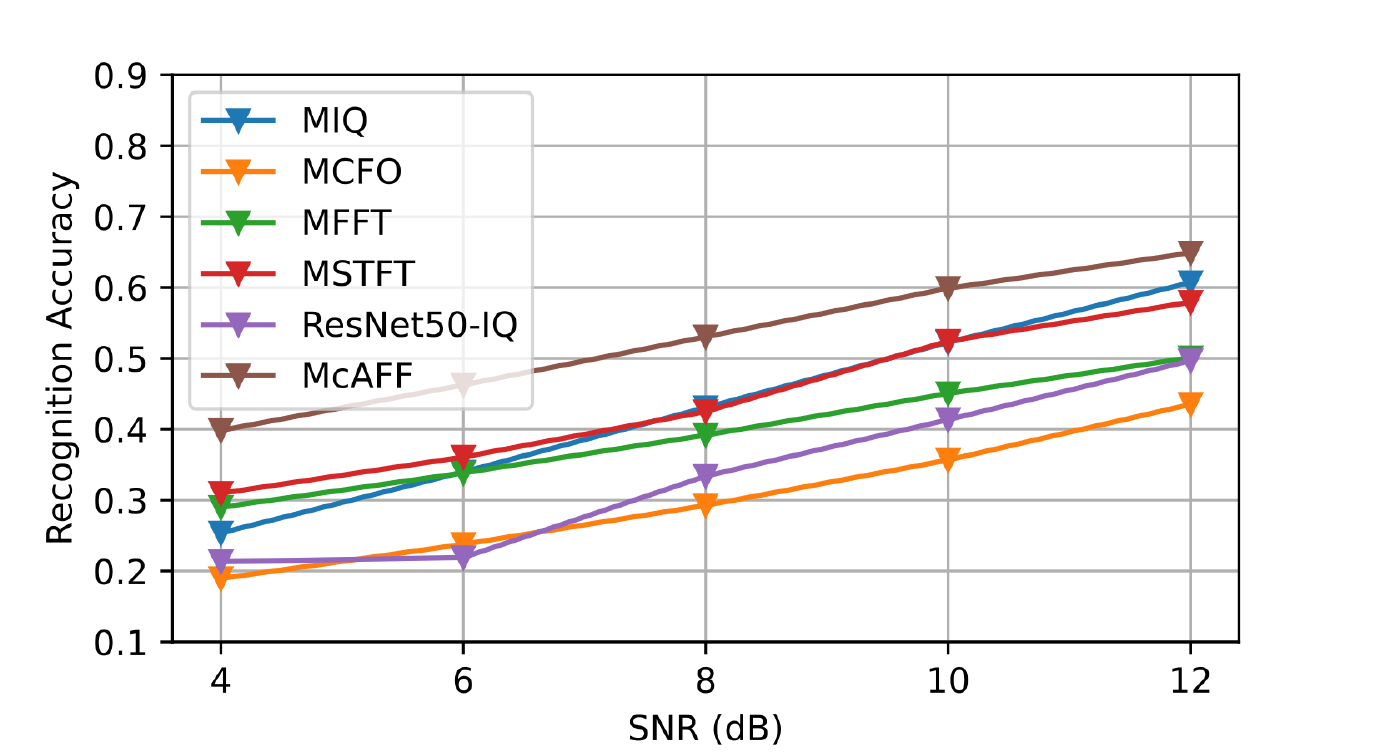}
%\caption{fig2}
\end{minipage}
}%
\subfigure[4\_train/4\_test.]{
\begin{minipage}[t]{0.5 \linewidth}
\centering
\includegraphics[width=3.5in]{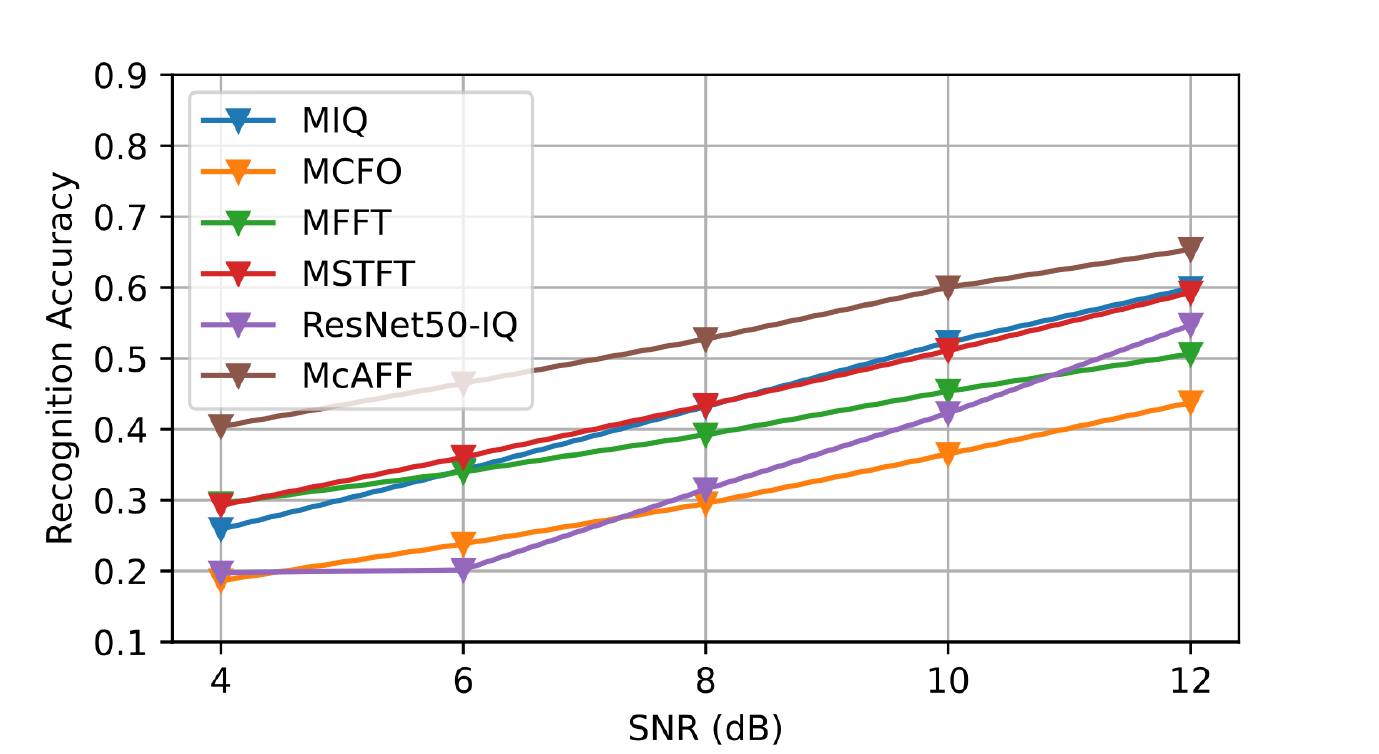}
%\caption{fig2}
\end{minipage}
}%
\centering
\caption{\textcolor{black}{RF fingerprint identificaiton performance versus SNR in the ``Train and Test Different Days'' scenario.}}
\label{fig:transRec}
\end{figure*}

\textbf{Impact of noise:} \textcolor{black}{The impact of noise on identification performance is shown in Fig.~\ref{fig:noise}. The additive noise is generated in a simulated way such that 14 different SNR levels from 4 dB to 30 dB are tested. It is shown  that the identification accuracy of all the RF fingerprinting methods increases with SNR. At high SNRs, the performance approaches that in Table~\ref{table:data_size_vs_acc}, which is based on almost clean WiFi signals collected in an anechoic chamber. The performance of a recent state-of-the-art: ResNet50-IQ\ \cite{al2020exposing} is also plotted in Fig.~\ref{fig:noise}. ResNet50-IQ directly used ResNet50\ \cite{7780459} as the signal identification network with IQ information as the input. It is clear that the proposed McAFF method significantly outperforms ResNet50-IQ, as well as the other single-channel RF fingerprinting methods. Closer observations show that MSTFT and MIQ have similar identification performance and outperform MFFT at all SNR levels. When the SNR is lower than 20 dB, ResNet50-IQ has better performance than both MCFO and MFFT, and the performance improvement of MFFT over MCFO is significant. When the SNR level is increased above 20 dB, MCFO starts to perform better than ResNet50-IQ and MFFT, and ResNet50-IQ shows similar performance as MFFT. Fig.~\ref{fig:transRec} further shows the RF fingerprint identification performance at different SNR levels when the training and test data are collected on different days. Compared with the single-channel RF fingerprinting models, the proposed method has the best identification performance at all SNR levels. The experimental results demonstrate that the proposed method that uses multi-channel attention fusion is robust against the noisy signals.}  

\textbf{Effect of feature attention:} \textcolor{black}{In Fig.~11, we compare the proposed McAFF model with a simplified version that concatenates multi-channel features without using the feature attention module in Fig.~\ref{fig:attention}. 
It is clear to see that the proposed McAFF model shows constantly improved identification performance at different SNRs. Fig.~11 demonstrates the effectiveness of the proposed shared attention module since it is able to dynamically control the weights of multi-channel neural features under various conditions.}

\begin{figure}[t!]
  \centering
  \includegraphics[width=\linewidth]{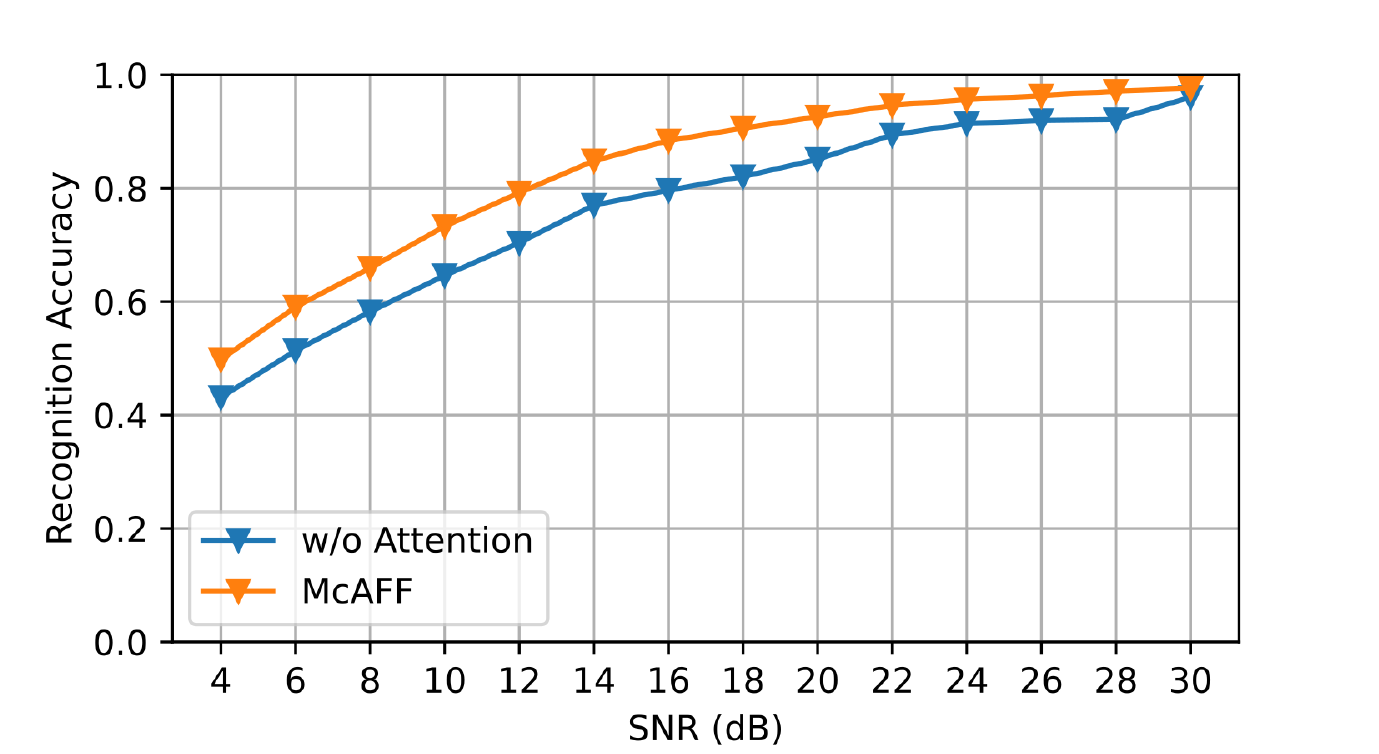}
  \caption{\textcolor{black}{Effect of feature attention. The training dataset size is $50\%N_{t}$.}}
  \label{fig:AFF}
\end{figure}

\section{Conclusion}
\label{con}
In this paper, we proposed an effective McAFF model for RF fingerprinting. The proposed model consists of two sequential parts: multi-channel feature fusion for feature learning and fusion, and signal identification for RF device identification. The multi-channel feature fusion part first computes the representations of the input WiFi signals through CFO estimation, FFT and STFT operations. Then, a shared attention module is used to adaptively control the weight of the neural features learned from the four channel inputs, including IQ samples, CFO, FFT and STFT coefficients. The model is trained to minimize the cross-entropy loss function. To evaluate the effectiveness of the proposed McAFF model for real-world RF signals, we designed a dataset (WFDI) captured from 72 WiFi end-devices over the course of 8 days in an anechoic chamber. We conducted an extensive evaluation of the impact of signal representation on RF fingerprint identification performance, and analyzed the effectiveness of the proposed shared attention module in terms of fingerprinting accuracy in scenarios where training and test data were collected on different days. In addition, we evaluated the identification performance of the proposed McAFF model on a simulated noisy dataset. Experimental results demonstrated that the proposed RF fingerprinting model with multi-channel feature fusion achieves much better identification performance on datasets of different sizes and different collection dates, and is more robust in noisy environments. \textcolor{black}{Learning to fuse multi-channel features is a straightforward approach to improving the design of a feature-based RF fingerprinting system. 
In this work, the RF signal acquisition system was placed in an anechoic chamber with simulated noise. Exploring and improving the proposed method for RF fingerprinting in a more dynamic communication environment is an interesting direction for future research.}

\bibliographystyle{IEEEtran}
\bibliography{references}

\end{document}